\documentclass[twocolumn,showpacs,preprintnumbers,amsmath,amssymb]{revtex4}

\usepackage{graphicx}
\usepackage{dcolumn}
\usepackage{bm}

\begin{document}

\preprint{APS/123-QED}

\title{Polarization dependence of emission spectra of multiexcitons in self-assembled quantum dots
}
\author{N.Y. Hwang}
\author{S.-R. Eric Yang}
\altaffiliation[corresponding author, ] {eyang@venus.korea.ac.kr}
\affiliation{Physics Department, Korea  University, Seoul Korea 
}


\begin{abstract}

We have investigated the polarization dependence of the emission spectra
of p-shell multiexcitons of a quantum dot when the single particle
level spacing is larger than the characteristic energy of the Coulomb interactions. We find that there are many degenerate multiexciton states. 
The emission intensities depend on the number of degenerate initial
and final states of the optical transitions. However, unlike the transition energies,
they are essentially independent of the strength of the Coulomb interactions.
In the presence of electron-hole symmetry the independence is exact.

\end{abstract}

\maketitle

\section{Introduction}

During recent years great progress has been made in the development of experimental tools for study
of optical properties of single dot structures\cite{Bry}:  Confocal microscopy and 
near field scanning microscopy have been particularly
useful.  
Self-assembled quantum dots(SAQDs) are promising as single photon 
generators\cite{Mi,Zhi} because of the high optical quality of the resulting dot structures.
Photons may  be generated from states formed by a  number of electrons and holes, i.e., 
multiexcitons\cite{marzin,faf,wyang,dekel1,dekel2,landin,toda,zre,Bry1,Bry2,Ike,Dek,Bar,Haw,Shum,Cor}.
Fine structures of excitons,  biexcitons, and triexcitons have been studied  
experimentally \cite{marzin,faf,wyang,dekel1,dekel2,landin,toda,zre}.
Exact numerical diagonalization and quantum Monte Carlo 
studies of emission spectra of multiexciton states have been performed 
for several different shapes of SAQDs\cite{Bry1,Bry2,Ike,Dek,Bar,Haw,Shum,Cor}.   
The emission spectra of Bose-Einstein condensed magnetoexcitons in the strong magnetic field limit have been also investigated
theoretically in SAQDs\cite{Yang}.

\begin{figure}[!hbt]
\begin{center}
\includegraphics[width = 0.3 \textwidth]{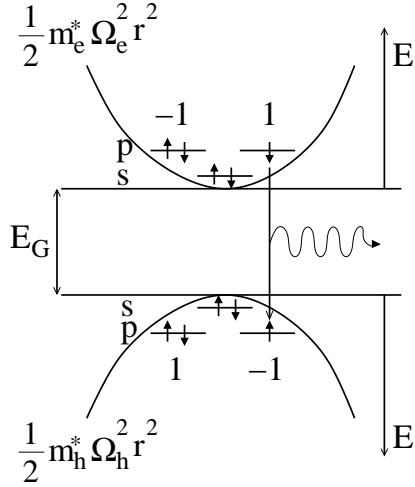}
\caption{Emission from the p-shells when N pairs of electrons and heavy holes occupy s- and p- orbitals of two-dimensional harmonic oscillators.
Note that both electron and hole 
energies are measured positive. The  bandgap $E_\mathrm{G}$
defined as the energy difference between the bottom of the electron harmonic potential and
the top of the hole harmonic potential. The numbers $1$ and $-1$ stand for the  angular momentum components along the axis perpendicular
to the two-dimensional layers.}
\label{fig:bandgap}
\end{center}
\end{figure}

\begin{figure}[!hbt]
\begin{center}
\includegraphics[width = 0.4 \textwidth]{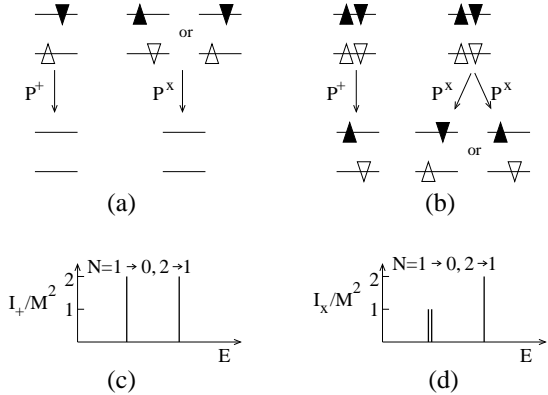}
\caption{Optical transitions between the s-shells. Filled (open) triangles represent electron (hole) spins.
(a) Possible transitions for circular ($P^+$) and linear ($P^x$) 
polarizations when $N=1\rightarrow0$.  (b) Possible transitions when $N=2\rightarrow1$.
(c) Schematic display of the emission spectra for right circular polarization.  (d) Emission spectra for linear  polarization.
}
\label{fig:N2.to.N1}
\end{center}
\end{figure}

However,   the  polarization dependence of the emission spectra of multiexcitons
in quantum dots is not well understood.
In one-, two-, and three-dimensional electron-hole plasmas  with parabolic bands the ratio between the emission 
intensities 
for linear and circular polarizations is constant since their optical strengths  
are $M^2$ and $2M^2$, see Table \ref{table:polarizationMat}.
(The constant $M$ is the dipole matrix element $-e \langle x |x|s \rangle$,
where $|x \rangle$ and $|s \rangle$ are the Bloch wave functions).
This is because the independent quasi-particle picture holds in these systems and each single particle state
is always doubly spin degenerate.
However, in zero-dimensional 
quantum dots this may not be the case since single particle levels are discrete and many body effects are strong,
see Figs. \ref{fig:bandgap} and  \ref{fig:N2.to.N1}.
It is unclear to what extent the independent quasi-particle picture holds.
A simple example demonstrates that there may be some  polarization dependence.
The  emission from the s-shells is simple to analyse and  
shows that  the emission spectra for linear and circular polarizations are
{\it not} proportional, see Fig. \ref{fig:N2.to.N1}.   This is because for $N=2\rightarrow1$ and 
linear polarization the initial state 
of the optical transitions
is non-degenerate
while the final states are doubly {\it degenerate}.
This means that the emission strength doubles from $M^2$ to $2M^2$.
We investigate polarization dependence of the emission spectra
of a quantum dot when a  photon is generated by the recombination of
an  electron-hole pair in the p-shells,
see Fig. \ref{fig:bandgap}. In this case the polarization dependence of the emission spectra is expected to be more
complicated since many body states can be strongly correlated.
Our results demonstrate that the optical strength  can depend sensitively on the polarization state of the emitted photon.
The optical strength can be enhanced significantly if there are several   degenerate multiexciton final states.
In addition the  optical strength can take different values depending on
which degenerate initial state of the optical transition is occupied.
The emission spectra are calculated using these optical strengths, and are given in Figs. \ref{fig:linear} and \ref{fig:circular}.
The emission intensities are almost independent of 
the strength of the Coulomb interactions in our model, although the transition energies do depend on the Coulomb interactions.
In the  presence of electron-hole symmetry the independence is exact.
Our results may be   useful since they could provide an important qualitative result that must 
be valid in a  more general approximation.

\begin{table}[!hbt]
\begin{tabular}{l|l|l|l|l}
\hline
Polarization & $x+i y$ & $x-i y$ & $x$ & $y$  \\
\hline
$|\mathrm e,1/2 \rangle \rightarrow |\mathrm h,-3/2 \rangle $ & $0$ & $\sqrt 2 M$ & $M$ & $i M$  \\
\hline
$|\mathrm e,-1/2 \rangle \rightarrow |\mathrm h,3/2 \rangle $ & $\sqrt 2 M$ & $0$ & $M$ & $-i M$  \\
\hline
\end{tabular}
\caption{Transition matrix elements between a  heavy hole and 
an electron for different polarizations.  
Note that a hole is the absence of a valence electron, and its spin takes the opposite value of the valence electron.
}
\label{table:polarizationMat}
\end{table}

\section{Model Hamiltonian}

For our purpose 
we may take a simple model, which  is illustrated in Fig. \ref{fig:bandgap}.  
In lens-shaped self-assembled quantum dots the conduction and valence band electrons 
can be approximated by those of harmonic oscillator wavefunctions.
Since in quantum dots only states near $\vec{k}=0$ are relevant we will neglect the heavy and
light hole mixing and concentrate only on the physics of the heavy hole.
The typical energy spacing 
between single particle energies is $20-40$meV, which is larger
than the strength of the Coulomb interaction of order $10$meV.
The energies of s-level and p-level  orbitals are denoted by $ \epsilon_{0}^\mathrm{e,h}$
and $\epsilon_{\pm1}^\mathrm{e,h}$.
Electron  wavefunctions are $u_\mathrm{c}(\vec{r})\chi(\pm\frac{1}{2})f_{\ell}(\vec{r})$
and heavy hole wavefunctions are
$ u_{\mathrm{v},\pm1}(\vec{r})\chi(\pm\frac{1}{2})g_{\ell}(\vec{r})$.
Here the conduction and heavy hole valence band 
Bloch wavefunctions are denoted by  $u_\mathrm{c}(\vec{r})$ and $u_{\mathrm{v},\pm1}(\vec{r})$.
In the following we will represent hole states $u_{\mathrm{v},1}(\vec{r})\chi(\frac{1}{2})g_{\ell}(\vec{r})$
and $u_{\mathrm{v},-1}(\vec{r})\chi(-\frac{1}{2})g_{\ell}(\vec{r})$ as having 
spin components $1/2$ and $-1/2$.
The two-dimensional  electron and hole effective mass wavefunctions are $f_{\ell}(\vec r_1)$ and $g_{\ell}(\vec r_2)$.
The quantities $a_\mathrm{e}^2=\frac{\hbar}{2 m_\mathrm{e}^* \Omega_\mathrm{e}}$ 
and $a_\mathrm{h}^2=\frac{\hbar}{2 m_\mathrm{h}^* \Omega_\mathrm{h}}$ are the characteristic length scales for electrons and holes.
The parameters $m_\mathrm{e}^*$ and $m_\mathrm{h}^*$
are the effective masses of electrons and holes,  
and $\Omega_\mathrm{e}$  and $ \Omega_\mathrm{h}$ are the strengths of the harmonic potentials.
Two-dimensional polar coordinates are $r_1$, $r_2$ and $\theta$.  We assume that the wavefunctions have a
width $d$ along the axis perpendicular to the two-dimensional plane.
This effect is 
included in the shape of the Coulomb interactions.
The {\it quasi-two-dimensional} Coulomb interaction has the form
$V(\vec r_1,\vec r_2)=\frac{e^2}{\varepsilon \sqrt{|\vec r_1-\vec r_2|^2+d^2}}$, where
$d$ is the width of the single particle wavefunctions along the z-axis ($\varepsilon$
is the background dielectric constant).
Single particle energies of electrons and holes are
$\varepsilon_{\ell}^\mathrm{e}=\hbar \Omega_\mathrm{e} (|\ell|+1)$ and
$\varepsilon_{\ell}^\mathrm{h}=\hbar \Omega_\mathrm{h} (|\ell|+1)$.
Note the hole energy is defined to be positive.  The  bandgap $E_\mathrm{G}$, 
defined as the energy difference between the bottom of the electron harmonic potential and 
the top of the hole harmonic potential, is set to zero.
In our model we assume that the single particle level spacing is
larger than or comparable to the
characteristic Coulomb energy, which is often valid in lens-shaped SAQDs\cite{karrai}. 
This allows us to ignore excitations from s to p levels and from p
to d levels/continuum states of two-dimensional layers \cite{foot2,karrai}.
We can thus  assume that   
the s-shell is completely filled and inert. 

The many body Hamiltonian $H$ for the electron-hole system can be written as a sum of
single particle terms $H_0$ and interaction terms $V_\mathrm{int}$, i.e., $H=H_0+V_\mathrm{int}$ with
$H_0=H_{\mathrm{e}}+H_{\mathrm{h}}$ and
$V_\mathrm{int}=H_{\mathrm{ee}}+H_{\mathrm{hh}}+H_{\mathrm{eh}}$,
where the single particle Hamiltonians
$H_{\mathrm{e}}=\sum_{\ell,\sigma } \epsilon_{\ell}^\mathrm{e} a_{\ell \sigma}^\dagger a_{\ell \sigma}$
and
$H_{\mathrm{h}}=\sum_{\ell,\sigma} \epsilon_{\ell}^\mathrm{h} b_{\ell \sigma}^\dagger b_{\ell \sigma}$.
The operator
$a^\dagger_{\ell,\sigma}$ and $b^\dagger_{\ell,\sigma}$ create an electron and a heavy hole 
($\sigma=\uparrow \textrm{or} \downarrow$).
Electron-electron, hole-hole, and electron-hole interaction terms are $H_\mathrm{ee}$, $H_\mathrm{hh}$, and $H_\mathrm{eh}$, respectively.
The electron Coulomb matrix elements are
$U_{\alpha_1 \alpha_2 \alpha_4 \alpha_3}^{\mathrm{ee}}=
\delta_{\sigma_1,\sigma_4}\delta_{\sigma_2,\sigma_3}\delta_{\ell_1+\ell_2,\ell_3+\ell_4}
\langle f_{\ell_1}f_{\ell_2}| V |f_{\ell_4}f_{\ell_3}\rangle  $ and hole  Coulomb matrix elements are
$U_{\beta_1 \beta_2 \beta_4 \beta_3}^{\mathrm{hh}}=\delta_{\sigma_1,\sigma_4}\delta_{\sigma_2,\sigma_3}
\delta_{\ell_1+\ell_2,\ell_3+\ell_4}
\langle g_{\ell_1}g_{\ell_2}| V |g_{\ell_4}g_{\ell_3}\rangle$
($\alpha$ and $\beta$ stand for $(\ell,\sigma)$).
The electron-hole matrix elements are 
$U_{\beta_1 \alpha_2 \beta_4 \alpha_3}^{\mathrm{eh}}=-\delta_{\sigma_1,\sigma_4}\delta_{\sigma_2,\sigma_3}
\delta_{\ell_1+\ell_2,\ell_3+\ell_4}
\langle g_{\ell_1}f_{\ell_2}| V |g_{\ell_4}f_{\ell_3}\rangle$.
Since the single particle energies are independent of spin we can write
$\varepsilon_{\ell, \sigma}^\mathrm{e} = \varepsilon_\ell^\mathrm{e}$.
When 
$\delta_{\sigma_1,\sigma_4} = \delta_{\sigma_2,\sigma_3 }=1$
we may write
$U_{\ell_1 \sigma_1,\ell_2 \sigma_2, \ell_4 \sigma_4, \ell_3 \sigma_3}^\mathrm{ee}
= V_{\ell_1,\ell_2,\ell_4,\ell_3}^\mathrm{ee}$. Similarly, simpler notations may be introduced
for $U^\mathrm{hh}$ and $U^\mathrm{eh}$.

\section{Multiexciton Hamiltonian matrix}

The total z-component of angular momentum  
$L_\mathrm{z}=L_{\mathrm{e},z}+L_{\mathrm{h},z}$ is conserved.
The electron and hole angular momenta, $L_{\mathrm{e},z}$ and $L_{\mathrm{h},z}$, need not be conserved  separately 
since $H_\mathrm{eh} $ can change them.
On the other hand the total z-component of spin for electrons  (holes), 
$S_{\mathrm{e},z}$  ($S_{\mathrm{h},z}$),
is conserved since $H_\mathrm{eh}$ cannot change $S_{\mathrm{e},z}$ or  $S_{\mathrm{h},z}$.
If $S_{\mathrm{e},z}$ is known then the electron spin quantum number $S_{\mathrm{e}}$ can be deduced: When  
$S_{\mathrm{e},z}=-1,0,\ \textrm{or}\ 1$ then $S_{\mathrm{e}}=1$, and 
when $S_{\mathrm{e},z}=-\frac{1}{2}\ \textrm{or}\ \frac{1}{2}$ then  $S_{\mathrm{e}}=\frac{1}{2}$.   In our model  $S_{\mathrm{e},z}$ 
cannot take other values than the ones just listed, see Fig. \ref{fig:levels}.
Similar results hold for hole spins.
We label the Hilbert subspaces  by 
$|L_{z},S_{\mathrm{e},z}, S_{\mathrm{h},z}\rangle$.
The resulting Hilbert subspaces are listed in the Fig. \ref{fig:levels} (see Appendix B).

The basis vectors in each Hilbert subspace may be chosen as 
the single Slater determinant states
$|\phi_i \rangle =
a^\dagger_{\alpha_N} \cdots a^\dagger_{\alpha_1} b^\dagger_{\beta_N} \cdots b^\dagger_{\beta_1}|0\rangle$, 
with $i=(\vec{\alpha_i},\vec{\beta_i})$ where        $\vec{\alpha_i}=(\alpha_1,\alpha_2,...,\alpha_N)$ and 
$\vec{\beta_i}=(\beta_1,\beta_2,...,\beta_N)$.
The $k$th eigenstate of $N$ electron-hole system 
is written as $|\Phi_k \rangle =\sum_i  A_i^k  |\phi_i \rangle  $.
The eigenstate $|\Phi_k \rangle $ and eigenvalue $E_k$ satisfy the following matrix equation
$\sum_j\langle \phi_i  |H|\phi_j \rangle  A_j^k = E_k A_i^k$.
The diagonal elements of the Hamiltonian matrix are 
\begin{eqnarray}
\langle \phi_i  |H|\phi_i \rangle=E_\mathrm{e}+E_\mathrm{h}+E_\mathrm{eh}.
\label{eq:diagonal}
\end{eqnarray}
The total energy of the electrons/holes can be determined from 
\begin{eqnarray}
E_p[n_\alpha]=\sum_\alpha \varepsilon_\alpha^p n_\alpha^p +\frac{1}{2}
\sum_{\alpha \beta}n_\alpha^p n_\beta^p U_{\alpha \beta}^{p},
\label{eq:diagonalelement}
\end{eqnarray}
with $U_{\alpha \beta}^{p}=U_{\alpha \beta \alpha \beta}^{pp}-U_{\alpha \beta \beta \alpha}^{pp}$,
where $p=\mathrm{e},\mathrm{h}$.
The electron and hole occupation numbers in $|\phi_i \rangle$ are denoted by $n_\alpha^\mathrm{e}$ 
and $n_\alpha^\mathrm{h}$.
The total electron-hole interaction energy is 
$E_\mathrm{eh}=\sum_{\alpha \beta} n_\alpha^\mathrm{e} n_\beta^\mathrm{h} U_{\alpha \beta \alpha \beta}^\mathrm{eh}$.
The off-diagonal elements are
\begin{eqnarray}
\langle \phi_i  |H|\phi_j \rangle=\langle \phi_i  |H_\mathrm{ee}|\phi_j \rangle+\langle \phi_i  |H_\mathrm{hh}|\phi_j \rangle
+\langle \phi_i  |H_\mathrm{eh}|\phi_j \rangle. \nonumber\\
\end{eqnarray}

\section{Eigenstates}

In our model  
it is possible to find analytically all the eigenstates in each Hilbert subspace.
This is because the maximum number of basis states of these Hilbert subspaces is at the most six and
the Hamiltonian matrices have simple forms.

\subsection{Uncorrelated single Slater determinant states and correlated states with two basis vectors}
In our model many Hilbert subspaces are one-dimensional.
They are  displayed in  Fig. \ref{fig:levels} and their energies are 
given by  Eq. (\ref{eq:diagonal}).
There are many two-dimensional Hilbert subspaces.  Their  Hamiltonian matrices are  symmetric
and    the diagonal elements are  the same:
\begin{eqnarray}
\left(
\begin{array}{cc}
\langle \phi_1 |H| \phi_1 \rangle & \langle \phi_1 |H| \phi_2 \rangle \\
\langle \phi_2 |H| \phi_1 \rangle & \langle \phi_2 |H| \phi_2 \rangle
\end{array}
\right)
=
\left(
\begin{array}{cc}
\gamma & \delta \\
\delta & \gamma
\end{array}
\right).
\end{eqnarray}
The  eigenvalues  of this matrix are
\begin{eqnarray}
E_1=\gamma+\delta,
\ E_2=\gamma-\delta,
\label{eq:twobytwo}
\end{eqnarray}
and the corresponding eigenvectors are
\begin{eqnarray}
|\Phi_1\rangle= \frac{1}{\sqrt 2} (|\phi_1\rangle +|\phi_2\rangle),\
|\Phi_2\rangle= \frac{1}{\sqrt 2} (|\phi_1\rangle -|\phi_2\rangle).
\end{eqnarray}
Note that these eigenvectors are {\it independent} of the Coulomb interactions.
The off-diagonal element $\delta=\langle \phi_1 |H| \phi_2 \rangle$ is non-zero  for 
$N=3,4,5$.

\subsection{Correlated states with six basis vectors}

In our model there is only one Hilbert subspace with six basis states:
the Hilbert subspace  $|L_{z}, S_{\mathrm{e},z}, S_{\mathrm{h},z}\rangle=|0,0,0\rangle$ for $N=4$.
The basis vectors are defined as follows:
$|\phi_1\rangle = a_{1,\uparrow}^\dagger a_{-1,\downarrow}^\dagger
b_{1,\uparrow}^\dagger b_{-1,\downarrow}^\dagger |S\rangle, 
|\phi_2\rangle = a_{1,\uparrow}^\dagger a_{-1,\downarrow}^\dagger
b_{1,\downarrow}^\dagger b_{-1,\uparrow}^\dagger |S\rangle,
|\phi_3\rangle = a_{1,\downarrow}^\dagger a_{-1,\uparrow}^\dagger
b_{1,\uparrow}^\dagger b_{-1,\downarrow}^\dagger |S\rangle, 
|\phi_4\rangle = a_{1,\downarrow}^\dagger a_{-1,\uparrow}^\dagger
b_{1,\downarrow}^\dagger b_{-1,\uparrow}^\dagger |S\rangle, 
|\phi_5\rangle = a_{-1,\downarrow}^\dagger a_{-1,\uparrow}^\dagger
b_{1,\downarrow}^\dagger b_{1,\uparrow}^\dagger |S\rangle, 
|\phi_6\rangle = a_{1,\downarrow}^\dagger a_{1,\uparrow}^\dagger
b_{-1,\downarrow}^\dagger b_{-1,\uparrow}^\dagger |S\rangle$,
where $|S\rangle=a_{0,\downarrow}^\dagger a_{0,\uparrow}^\dagger 
b_{0,\downarrow}^\dagger b_{0,\uparrow}^\dagger |0\rangle$ represents the filled s-shells. 
Note that $|\phi_5\rangle$ and $|\phi_6\rangle$ contain doubly occupied p-orbitals.
The diagonal  matrix elements are $A=\langle \phi_k |H| \phi_k \rangle$ and are
given by Eq. (\ref{eq:diagonal}).
If we define
$B=-V_{-1,1,1,-1}^\mathrm{ee},C=-V_{-1,1,1,-1}^\mathrm{hh}$ and $D=V_{-1,1,1,-1}^\mathrm{eh}$,
the Hamiltonian matrix is given by
\begin{eqnarray}
&H&=\left(
\begin{array}{cccccc}
A & C & B & 0 & D & D \\
C & A & 0 & B & -D & -D \\
B & 0 & A & C & -D & -D \\
0 & B & C & A & D & D \\
D & -D & -D & D & A & 0 \\
D & -D & -D & D & 0 & A 
\end{array}
\right),
\end{eqnarray}
with the  eigenvalues
$E_1=A+B+C$,
$E_2=\frac{1}{2}(2A-B-C-\sqrt{B^2+2BC+C^2+32D^2})$,
$E_3=A+B-C$,
$E_4=A$, 
$E_5=A-B+C$ and
$E_6=\frac{1}{2}(2A-B-C+\sqrt{B^2+2BC+C^2+32D^2})$.
The eigenvectors are
\begin{eqnarray}
&&|\Phi_1\rangle= \frac{1}{2} (|\phi_1\rangle +|\phi_2\rangle +|\phi_3\rangle +|\phi_4\rangle),\nonumber \\
&&|\Phi_2\rangle= a_1 |\phi_1\rangle +a_2 |\phi_2\rangle +a_3 |\phi_3\rangle +a_4 |\phi_4\rangle
+a_5 |\phi_5\rangle +a_6 |\phi_6\rangle,\nonumber\\
&&|\Phi_3\rangle= \frac{1}{2} (-|\phi_1\rangle +|\phi_2\rangle -|\phi_3\rangle +|\phi_4\rangle),\nonumber \\
&&|\Phi_4\rangle= \frac{1}{\sqrt 2} (-|\phi_5\rangle +|\phi_6\rangle),\nonumber\\
&&|\Phi_5\rangle= \frac{1}{2} (-|\phi_1\rangle -|\phi_2\rangle +|\phi_3\rangle +|\phi_4\rangle),\nonumber\\
&&|\Phi_6\rangle= b_1 |\phi_1\rangle +b_2 |\phi_2\rangle +b_3 |\phi_3\rangle +b_4 |\phi_4\rangle
+b_5 |\phi_5\rangle +b_6 |\phi_6\rangle.\nonumber\\
\label{eq:evec}
\end{eqnarray}
Note again that $|\Phi_1\rangle,|\Phi_3\rangle,|\Phi_4\rangle,$ 
and $|\Phi_5\rangle$ are {\it independent} of the Coulomb interactions and model parameters.
On the other hand , the expansion coefficients of $|\Phi_2\rangle$   and $|\Phi_6\rangle$ 
do depend on them.
The expansion coefficients can be obtained analytically but
their expressions are rather complicated and lengthy.  
The lowest energy state  is $|\Phi_1\rangle$ since $B<0,C<0$.   In this state both electrons and holes are
in a spin triplet  state  $ S_e=1$ and $S_h=1$ with  $(S_{e,z},S_{h,z})=(0,0)$.

\subsection{Number of groundstate degeneracy}

Our investigation shows  groundstates have $L_z=0$, see Fig. \ref{fig:levels}.  When $N=6,5,4,3,2$ we
see from Fig. \ref{fig:levels} that the number of degenerate states with $L_z=0$  are $1,   4,   9,  4,   1$.
Let us examine why the degeneracy is 9 for  $N=4$.
From  symmetry considerations one can predict the number of groundstate degeneracy.
Confirming this number of degeneracy provides a powerful {\it check} of the correctness of our analytic calculation
of eigenenergies.
We find that   the lowest energy states are  the ones with 
electrons and holes  in triplet states.
The electron and hole spin numbers are thus $S_\mathrm e=1$ and $S_\mathrm h=1$.
This implies that the possible values of $(S_{\mathrm e,z},S_{\mathrm h,z})$  are $3\times3=9$. 
Since the interactions are spin invariant these states are degenerate.
Our calculation shows that the groundstates of the following nine Hilbert subspaces 
$| 0,1,1 \rangle$,
$| 0,0,1 \rangle$,
$| 0,0,-1 \rangle$,
$| 0,-1,-1 \rangle$,
$| 0,-1,1 \rangle$,
$| 0,1,-1 \rangle$,
$| 0,-1,0 \rangle$,
$| 0,1,0 \rangle$, and 
$|0,0,0 \rangle$ are degenerate (see Fig. \ref{fig:levels}). 
Other degeneracies may be shown in a similar manner.

\section{Optical strength and emission spectra}
\begin{table}[!hbt]
\begin{tabular}{l|c|c}
\hline
Transition types & Linear pol. & Circular pol.\\
 \hline
Between single and two-basis states
 & $2M^2$ & $4M^2$\\
 \hline
Between two-basis states
 & $M^2$ & $2M^2$\\
 \hline
Between two- and six-basis states
 & $\frac{1}{2}M^2$ & $M^2$\\
 \hline
\end{tabular}
\caption{In our model most optical dipole strengths  can take only
four different values, $\frac{1}{2}M^2, M^2, 2M^2\ \textrm{and}\ 4M^2$.  
The oscillator strengths of circular polarization are twice those of linear polarization,
just like in the case of single particle transitions.} 
\label{table:values}
\end{table}
\begin{figure}[!hbt]
\begin{center}
\includegraphics[width = 0.4 \textwidth]{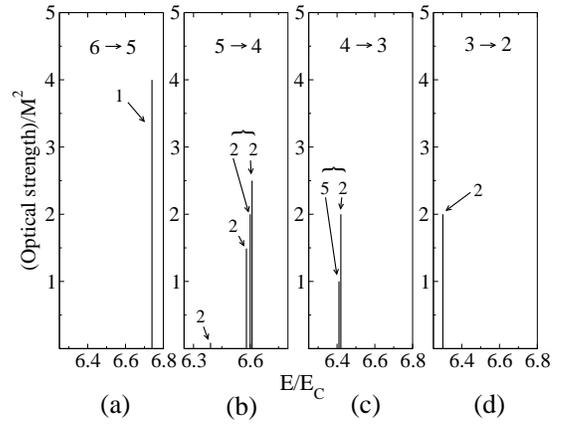}
\caption{
Emission spectra for linear polarization $P^x$.
Closely spaced vertical lines under the curly bracket have all the {\it same} transition energy.
The numbers above the arrows  denote the number of different initial states  giving rise
to the same transition energy. The change in the number of electron-hole pairs $N \rightarrow N-1$ 
is indicated in the upper part of each panel.
}
\label{fig:linear}
\end{center}
\end{figure}

\begin{figure}[!hbt]
\begin{center}
\includegraphics[width = 0.4 \textwidth]{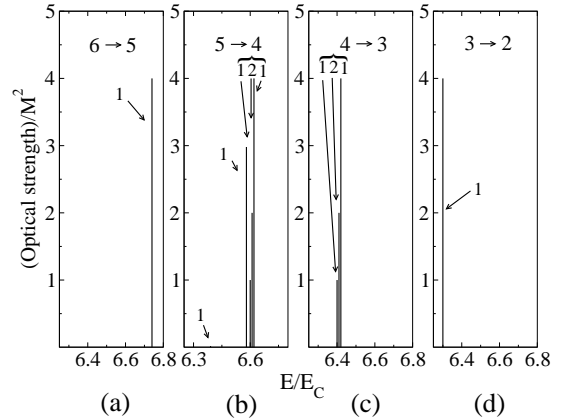}
\caption{
Emission spectra for circular polarization $P^+$.
}
\label{fig:circular}
\end{center}
\end{figure}

In this paper we consider 
linearly, right circularly, and left circularly polarized photons.   
We assume fast energy relaxation so that only emission from ground states of $N$ exciton pairs are needed.
The emission spectrum is given by 
\begin{eqnarray}
I(\omega)=\sum_f |\langle \Phi_f |\hat L | \Phi_\mathrm{G} \rangle|^2 \delta(E_i-E_f-\omega),
\end{eqnarray}
where the luminescence operator is
\begin{eqnarray}
\hat L = \sum_{\ell,\sigma} \langle -\ell,-\sigma |\hat \varepsilon \cdot \vec p | \ell,\sigma \rangle 
b_{-\ell,-\sigma} a_{\ell,\sigma}.
\end{eqnarray}
An electron and a hole can only recombine when they have opposite values of the z-component of angular momenta
and spins.  This leads to the  {\it selection rules}
$\Delta L_z=0$ and $\Delta S_z=0$.
The  optical dipole strength from a  ground state of N-exciton system, $\Phi_\mathrm{G}(N)$, to a ground/excited
state of (N-1)-exciton system, $\Phi_f(N-1)$, is  
$ |\langle \Phi_f |\hat L | \Phi_\mathrm{G} \rangle|^2$.  The possible values of the optical dipole strength
are displayed in Table \ref{table:values}.
In our model there are only three types of dipole transitions:
(a) transitions between single and two-basis states, 
(b) transitions   between two-basis states,
and (c) transitions  between two- and six-basis states.
We have computed  emission spectra for polarization along the x-axis  ($P^x$), left circular polarization ($P^{-}$), and 
right circular polarization ($P^{+}$).   They are  
shown in Figs. \ref{fig:linear} and \ref{fig:circular}.  
The following parameters are  used: $a_\mathrm{h}=0.46 a_\mathrm{e}$, $d=0.8 a_\mathrm{e}$, 
$\hbar \Omega_\mathrm e  = 2 E_\mathrm C$, and $\Omega_\mathrm h=0.7 \Omega_\mathrm e$. 
In our calculation we measure  energy with respect to  the Coulomb energy  $E_\mathrm C=\frac{e^2}{\epsilon a_\mathrm{e}}$. 
Note that the transition energy is actually $E_i-E_f+E_\mathrm{G}$, but  
we will drop the bandgap energy $E_\mathrm{G}$ for convenience hereafter.
The emission spectra for right and left circular polarizations are identical.
Note that  only one transition energy is possible when $N= 6 \rightarrow 5$, $4 \rightarrow 3$, and
$3 \rightarrow 2$.   This is 
because transitions from the ground state of $N=6$ to the excited states of $N=5$,
and from the ground state of $N=4$ to the excited states of $N=3$ are optically forbidden or vanish.
On the other hand, for $N=5 \rightarrow 4$ three transition energies are possible.
One can show that the excited states $|\Phi_3\rangle$, $|\Phi_4\rangle$ and $|\Phi_5\rangle$ 
of the Hilbert subspace $|0,0,0\rangle$ for $N=4$ do not contribute to optical strengths   (see Appendix A).
Only $|\Phi_1\rangle$, $|\Phi_2\rangle$ and $|\Phi_6\rangle$ contribute.

\begin{widetext}
\begin{figure}[hbt]
\begin{center}
\includegraphics[angle=90, height = 0.9 \textheight]{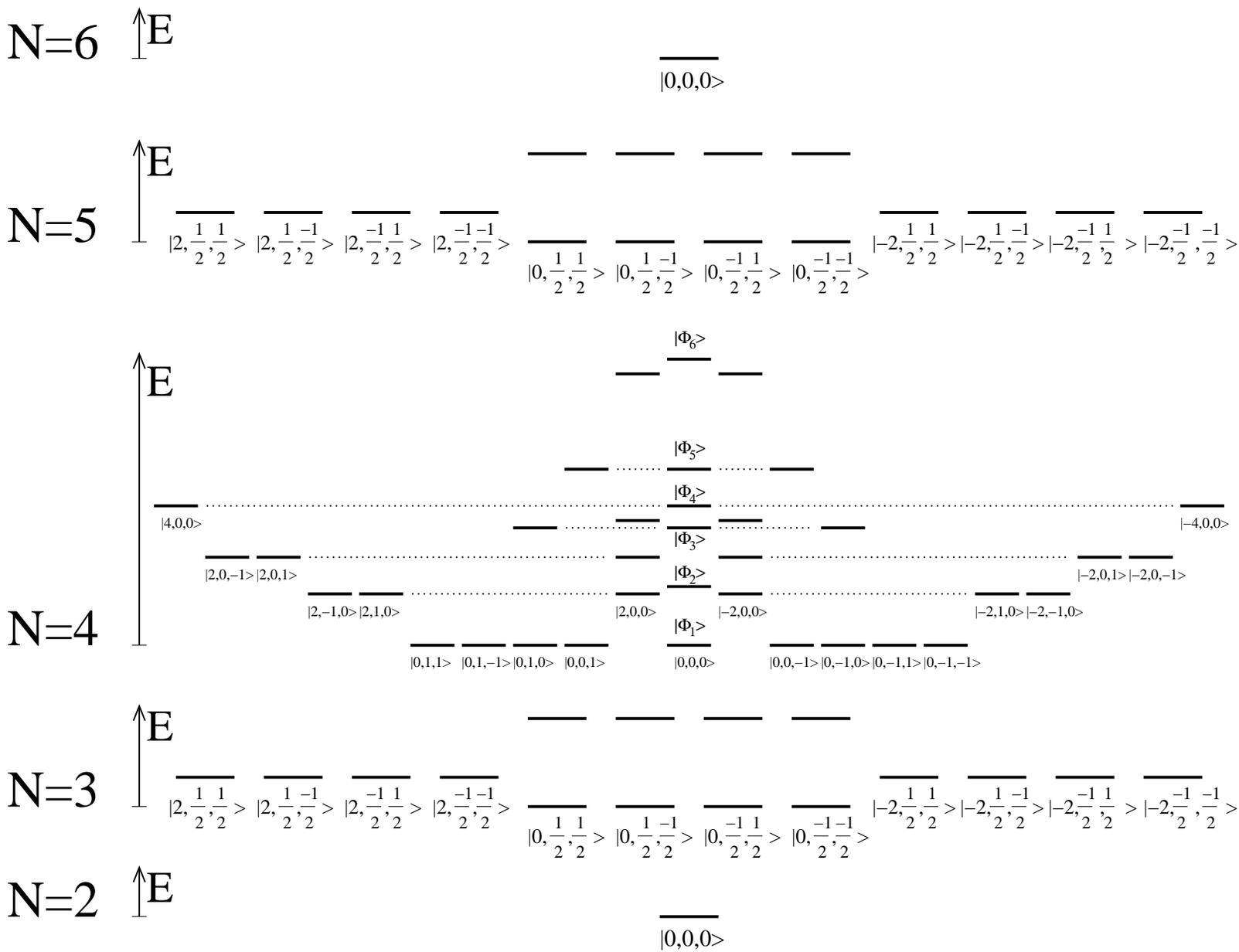}
\caption{Energy levels of each  Hilbert subspace $|L_z,S_{\mathrm{e},z},S_{\mathrm{h},z} \rangle$.
The dimension of each subspace is equal to the number of energy levels displayed.
For $N=4$ the states in the four dimensional Hilbert subspaces  $|2,0,0 \rangle$, $|-2,0,0 \rangle$ are optically inactive.} 
\label{fig:levels}
\end{center}
\end{figure}
\end{widetext}

Let us  discuss the emission spectra of $5\rightarrow4 $  for
$P^x$.   
The oscillator strength of  the lowest energy peak $6.39 E_C$ is $0.0088M^2$.  It originates from the transitions
$|0,\frac{1}{2},-\frac{1}{2}\rangle_1 \textrm{ or }|0,-\frac{1}{2},\frac{1}{2}\rangle_1
\rightarrow |\Phi_6\rangle$, see Fig. \ref{fig:levels} (These are defined as type A transitions).
The {\it subscript} $1$ in $|0,\frac{1}{2},-\frac{1}{2}\rangle_1 $ indicates that it is the lowest energy eigenstate.
This oscillator strength
is negligible compared to those of the others.
The oscillator strength of  the  next lowest  transition energy  $6.58 E_C$ is  $1.4912M^2$.
It originates from the transitions
$|0,\frac{1}{2},-\frac{1}{2}\rangle_1 \textrm{ or }|0,-\frac{1}{2},\frac{1}{2}\rangle_1 
\rightarrow  |\Phi_2\rangle$ (These are defined as type B transitions).
The dependence of the strength of this  peak on  the ratio $\Omega_\mathrm{h}/\Omega_\mathrm{e}$
is negligible: as it is reduced from $0.7$ to $0.1$ the oscillator strength 
changes smoothly from $1.4912M^2$ to $1.5000M^2$ (the change is only about $1\%$).
The dependence on  the Coulomb interactions, i.e., on the ratio $d/{a_\mathrm e}$ 
is also negligible: as it is reduced  from $0.8$ to $0.2$ 
the oscillator strength reduces continuously from $1.4912M^2$ to $1.4780M^2$.
As these parameters change the sum of the two oscillator strengths of transitions  A and B remains a constant equal to  $3/2M^2$.
This is {\it evidence} that our calculations are correct.  
The highest peak with the energy $6.61 E_C$ has the optical strength $5/2 M^2$ and originates from the transitions
$|0,\frac{1}{2},-\frac{1}{2}\rangle_1
\rightarrow|0,1,-1\rangle_1\textrm{ and }|\Phi_1\rangle,\textrm{ or }
|0,-\frac{1}{2},\frac{1}{2}\rangle_1
\rightarrow|0,-1,1\rangle_1\textrm{ and }|\Phi_1\rangle$,  see Fig. \ref{fig:levels}.
In these transitions there are two possible final states with the same energy,
and the oscillator strength is the sum of the two contributions.
The other  peak  at the same transition energy with  the strength $2M^2$ originates from the transitions
$|0,\frac{1}{2},\frac{1}{2}\rangle_1
\rightarrow|0,1,0\rangle_1\textrm{ and }|0,0,1\rangle_1,\textrm{ or }
|0,-\frac{1}{2},-\frac{1}{2}\rangle_1
\rightarrow|0,-1,0\rangle_1\textrm{ and }|0,0,-1\rangle_1$.
Note that some transition energies 
can have more than one optical strengths. For example, for  $N=5\rightarrow4$  at the transition energy $6.61E_C$
two values of optical strengths $2M^2$
and $5/2M^2$ are possible.
In this case there are two different initial states giving rise to  the same value of the  optical strength $2M^2$.  
The same is also true for $5/2M^2$, see Fig. \ref{fig:linear}. 

Now we discuss the emission spectra for $P^{+}$ for the case $N=5 \rightarrow 4$.
The oscillator strength of  the lowest energy peak $6.39 E_C$ is $0.0176M^2$.  It originates from the transition 
$|0,-\frac{1}{2},\frac{1}{2}\rangle_1 
\rightarrow |\Phi_6\rangle$, see Fig. \ref{fig:levels} (We  define it as type C transition).
This oscillator strength
is negligible compared to those of the others.
The oscillator strength of  the  next lowest  energy peak $6.58 E_C$ is  $2.9824M^2$.
It originates from the transitions
$|0,-\frac{1}{2},\frac{1}{2}\rangle_1
\rightarrow  |\Phi_2\rangle$ (We  define it  as type D transition).
Again as the physical  parameters change the sum of the two oscillator strengths of transitions  C and D remains a constant equal to  $3M^2$.
Note that some transition energies
can have more than one optical strengths. For example, for  $N=5\rightarrow4$  at the transition energy $6.61E_C$
three values of optical strengths $M^2$, $2M^2$ and $4M^2$ are possible.
The highest peak $6.61 E_C$ with the strength $4M^2$ originates from the transitions
$|0,\frac{1}{2},-\frac{1}{2}\rangle_1
\rightarrow|0,1,-1\rangle_1$.
The  peak $6.61 E_C$ with  the strength $2M^2$ originates from the transitions
$|0,\frac{1}{2},\frac{1}{2}\rangle_1
\rightarrow|0,1,0\rangle_1, \textrm{ or }
|0,-\frac{1}{2},-\frac{1}{2}\rangle_1
\rightarrow|0,0,-1\rangle_1$.
The  peak $6.61 E_C$ with  the strength $M^2$ originates from the transitions
$|0,-\frac{1}{2},\frac{1}{2}\rangle_1 
\rightarrow|\Phi_1\rangle$.
There are thus two different initial states giving rise to  the same value of the  optical strength $2M^2$,
and there are only one initial states giving rise to the  optical strengths $M^2$  and $4M^2$.

We have verified numerically that when electrons and holes have exactly the same properties,
i.e., when {\it electron-hole symmetry} is present, the dependence of the oscillator strengths for the transitions A, B, C and D on the Coulomb interactions
disappears completely. The oscillator strengths of these transitions become, respectively, $0$, $\frac{3}{2}$, $0$, and  $3$.

\section{Discussions}

Our main results may be attributed to the presence
of degenerate states and the strongly correlated nature
of multiexciton states. In the non-interacting model, spin
properties of electronic states do not change as a function of
energy. In the interacting case, spin properties of electronic
states do change as a function of energy. At different transition
energies spin properties of optically active initial and final states
are different. Which initial and final states are connected by the
optical selection rules will depend on the polarization of the
emitted photon. Several factors contribute to the quantitative
aspects of the polarization dependence of the emission spectra.
As illustrated in Fig. \ref{fig:N2.to.N1} the presence of degenerate final states
can enhance the oscillator strengths. The presence of degenerate
initial states can lead to the presence of degenerate transitions
energies. The correlated nature of some eigenstates influence
the value of their oscillator strengths and can lead to a
cancellation of the optical strength.

Our approximate calculation shows that there is a significant
polarization dependence. For more realistic SAQDs our results
may be used as a starting point of a perturbative treatment of the
electron-hole exchange interaction, the coupling between s and
p and between p and d levels, coupling between p-levels and the
continuum states of the two-dimensional layers, heavy and light
hole coupling, and elliptic dot confinement potential \cite{Soo}. We
expect in these cases that the emission intensities will depend
on the Coulomb interactions. However, as long as the single
particle level spacing is larger than the characteristic Coulomb
energy the effect would be minor.

\begin{acknowledgments}
This work was  supported by grant No. R01-2005-000-10352-0 from the Basic Research Program
of the Korea Science and Engineering
Foundation,
by Quantum Functional Semiconductor Research Center (QSRC) at Dongguk University
of the Korea Science and Engineering
Foundation 
and by Pure Basic Research Group project of Korea Research Foundation under Grant No. C00054.
This work was supported by The Second Brain 21 Project.
\end{acknowledgments}

\begin{references}
\bibitem{foot2}
In Ref. \cite{Haw} the couplings from the p-shell to  d-shell and  to the continuum states of  the two-dimensional layers are ignored. 
However, if  excitations from the s- to p-shells are significant then so are 
excitations from the p- to d-shells since
the probabilities for these excitations have the same order of magnitude.
If the barrier height of the quantum dots is small d-shells may not exist.  Even in this case
excitations  from the p-shell to the continuum states of the two-dimensional
layers must be included,  
see Ref. \cite{karrai}.
\end{references}
\end{document}